\documentstyle[12pt,epsfig]{article}

\title{Local log-law of the wall: numerical evidences and reasons} 
\author{S.~Besio, A.~Mazzino$^{*}$ and C.F.~Ratto\\
\small{ INFM -
Dipartimento di Fisica, Universit\`a di Genova, I--16146 
Genova, Italy.}}
\begin{document}
\maketitle
\date{}
\vspace*{-0.6cm}
\footnote{*~Corresponding author: mazzino@fisica.unige.it~~
tel. \& fax: 0039-010-3536354}
\begin{abstract}
Numerical studies performed with a primitive equation model 
on two-dimensional sinusoidal hills show that the local 
velocity profiles behave logarithmically to a very good approximation, 
from a distance from the surface of the order of the maximum hill height
almost up to the top of the boundary
layer. This behavior is well known for flows above homogeneous and 
flat topographies (``law-of-the-wall'') and, more recently,
investigated with respect to the large-scale (``asymptotic'') averaged
flows above complex topography.
Furthermore, this new-found local 
generalized law-of-the-wall involves effective parameters showing
a smooth dependence on the position along the 
underlying topography.
This dependence is similar to the topography itself, 
while this property does not absolutely hold for the underlying flow,
nearest to the hill surface.

\end{abstract}

Near-wall-turbulence

\vspace{4mm}

The impact of terrain features has been recognized 
(see, e.g., \cite{T86,W92}) to be
crucial for the correct prediction of the atmosphere general
circulation.
This is mainly due to the key role played by the surface features
in extracting momentum from the atmosphere, either through the
differential pressure across the object or the vertical propagation of
internal gravity waves initiated by the flow over the mountains.
The problem for the weather prediction
is that most of the surface disturbances 
are much smaller than what can be resolved by
any current operational numerical model, and yet these disturbances
can still have a considerable impact on the transfer of momentum from
the atmosphere to the surface. For this reason the problem concerning 
how to parameterize the small-scale orographic effects on the
large-scale dynamics attracts much attention
\cite{T81,E87,TSM89,WM93,EZ94}. 
An important result in this direction concerns 
the generalization to the case of complex terrain
of the well-known ``law-of-the-wall'' (see, e.g.,  \cite{YM})
valid for a neutral homogeneous turbulent boundary layer
above flat terrain:
\begin{equation}
U (z) =
\frac{u_{\star}}{k} 
\ln \left( \frac{z}{z_0} \right )\;\;\; ,
\label{logen}
\end{equation}
where $U(z)$ is the wind profile averaged over a time much larger
than the characteristic times of turbulence, $k$ is the Von K\`arm\`an's 
constant, $z$ is the distance from the surface,
$z_0$ and $u_{\star}$
are the ``roughness length'' and the ``friction velocity'',
respectively.
Just very recently 
the above logarithmic law-of-the-wall  for the mean velocity profile 
has received a rigorous analytical prove.  This has been obtained 
in Refs.~\cite{N99,NKD99}
by combining the ``rapid distortion theory'' and the averaged
Reynolds stress description of the mean flow.

It is probably worthwhile to recall (see, e.g., Ref.~\cite{YM}) 
that $z_0$, a quantity 
characteristic of the surface itself, is related to the height of
surface protrusions, while 
$u_{\star}$, a property of the flow, is proportional to the
turbulent fluxes of momentum along 
the vertical (see again Ref.~\cite{YM}). 

It has been actually found, as a result of numerical simulations (see,
e.g.,  Ref.~\cite{WM93})
and observations  (both in wind tunnel, see, e.g., 
Refs.~\cite{GTD96}, and in nature, see, e.g.,
Ref.~\cite{Hignett94}), that,
in a neutral homogeneous boundary layer over hilly terrain, 
the logarithmic shape is restored to a very reasonable approximation,
from a distance from the surface of the order of
the maximum hill height almost up to the top of the boundary, 
for $\langle U \rangle$, i.e. 
the velocity 
profile averaged over an area of linear dimensions
much larger than the typical scale on which the topography varies, 
a regime which we would  like to define 
as ``asymptotic'', in analogy with another topic we are going to
discuss at the end of this Letter.
More precisely, Eq.~(1) still holds, with 
$U(z)$ replaced by $\langle U \rangle (z)$, where $z$ is still the
distance from the surface,
$u_*$ is replaced by 
the ``effective friction velocity'', ${\sf u}_{\star}^{\mbox{\tiny eff}}$,
and 
$z_0$ is replaced by the ``effective roughness length'',
${\sf z}_0^{\mbox{\tiny eff}}$. 
Notice that the values of these effective parameters
are considerably larger than those of the corresponding ones
in the absence of any hill.

It is a matter of fact that one can be 
interested in the dynamics of the flow over intermediate scales 
between the two considered, i.e. at scales comparable 
with those on which the terrain varies, a regime which we like to refer
to as ``pre-asymptotic''. For this regime,
an interesting and natural question arises about the presence 
of some form of structural similarity 
and thus on the existence of logarithmic laws with
effective parameters through which the dynamics can be described.
If that is so, the investigation of the relation 
(if any) between the asymptotic and the pre-asymptotic parameters
should be an interesting issue to be investigated.
The above points, which are up to now largely unexplored, are the main 
concern of the present Letter.

To investigate the above 
conjectures, we have considered in detail the case-studies 
analyzed by Wood and Mason in Ref.~\cite{WM93} (hereafter, WM93
data-set), 
even if we have drawn similar conclusions
analyzing experiments both in a wind tunnel (see
Ref.~\cite{GTD96}) and in nature (see, e.g.,  Ref.~\cite{Hignett94})).
In fact,  the attention of
Wood and Mason, as well as that of any
other author involved with this topic -- at least as far as we know --
was focused on the areally averaged velocity field 
(over scales much greater than those on which topography varies)
and not on the scales we referred to as the pre-asymptotic regime.
 
The part of WM93 data-set we have considered 
consists of velocity fields obtained from
numerical simulations performed 
over three different two-dimensional topographies whose shapes are
described by the following expression:
\begin{equation}
h(x,y)=H sin^{2}\left(\frac{x\pi}{\lambda}\right)\;\;\; .
\label{collina}
\end{equation}
In all modeled cases, $\lambda =1000\; m$, while
the roughness length, $z_0$, is $0.1\;m$; 
three different surface configurations have been considered in the 
present study:
$H=250\;m$ (hereafter hill H250), $H=100\;m$ (H100) and $H=20\;m$ (H20).

The flow is driven by a constant pressure gradient corresponding
to a geostrophic wind speed of 10 $m/s$ in the $x$-direction.
The model was initialized with the ''unperturbed'' profile, i.e.
the wind profile
relative to a corresponding flat surface, shown as a dashed line in
Fig.~1, the flattening of $U(z)$ at values of $z$ higher than $\sim 1000
\; m$ being due to the 
geostrophic balance occurring at the end of the Ekman layer 
(see, e.g., Ref.~\cite{H79}) where the surface drag is vanishing.

The Wood and Mason's numerical model 
uses the Boussinesq approximation to the
ensemble-averaged Navier--Stokes equations,
together with the equation of mass conservation and
a $1\frac{1}{2}$-order closure scheme through which
small-scale (subgrid) motions are described. 
More precisely, the Reynolds'
stress is expressed in the form $\tau_{ij}=\nu S_{ij}$ where
$S_{ij}=(\partial_i u_j+\partial_j u_i)$ and the eddy viscosity $\nu$ is
modeled as $\nu=L_m^2S$, with $S=1/2\;S_{ij}S_{ij}$ and $L_m$ the
mixing length, obtained by solving the equation for the turbulent kinetic
energy (for major details, see Ref.~\cite{WM93} and the references
therein).

Equations are solved by standard finite difference methods 
with a grid spacing of approximately $50\;m$.
Periodic boundary conditions in both horizontal directions are also
imposed. A more extensive description of the numerical model 
and of its evaluation against field experiments can be found in   
Refs.~\cite{W92,WM93} 
and in other papers of the same research group.

Our remark concerning the analysis of WM93 data-set, is that 
the generalized law of the wall observed in Ref.~\cite{WM93} 
for $\langle U\rangle(z)$
is certainly present, but a more intrinsic logarithmic shape
\begin{equation}
U (x,z) =
\frac{u_{\star}^{\mbox{\tiny eff}}(x)}{k} 
\ln \left( \frac{z}{z_0^{\mbox{\tiny eff}}(x)}\right )\qquad\mbox{for}\qquad
z>z_{min}\sim H
\;\;\; ,
\label{generalizzo}
\end{equation}
where the effective parameters now show a dependence on the
horizontal position $x$,
actually takes place at smaller scales than those focused by the
authors. This
fact can be easily checked from the results shown in Fig.~1, where 
\setcounter{figure}{0}
\begin{figure}[ht]
\vfill \begin{minipage}{.4\linewidth}
\begin{center}
\vspace{0.0cm}
\mbox{\psfig{figure=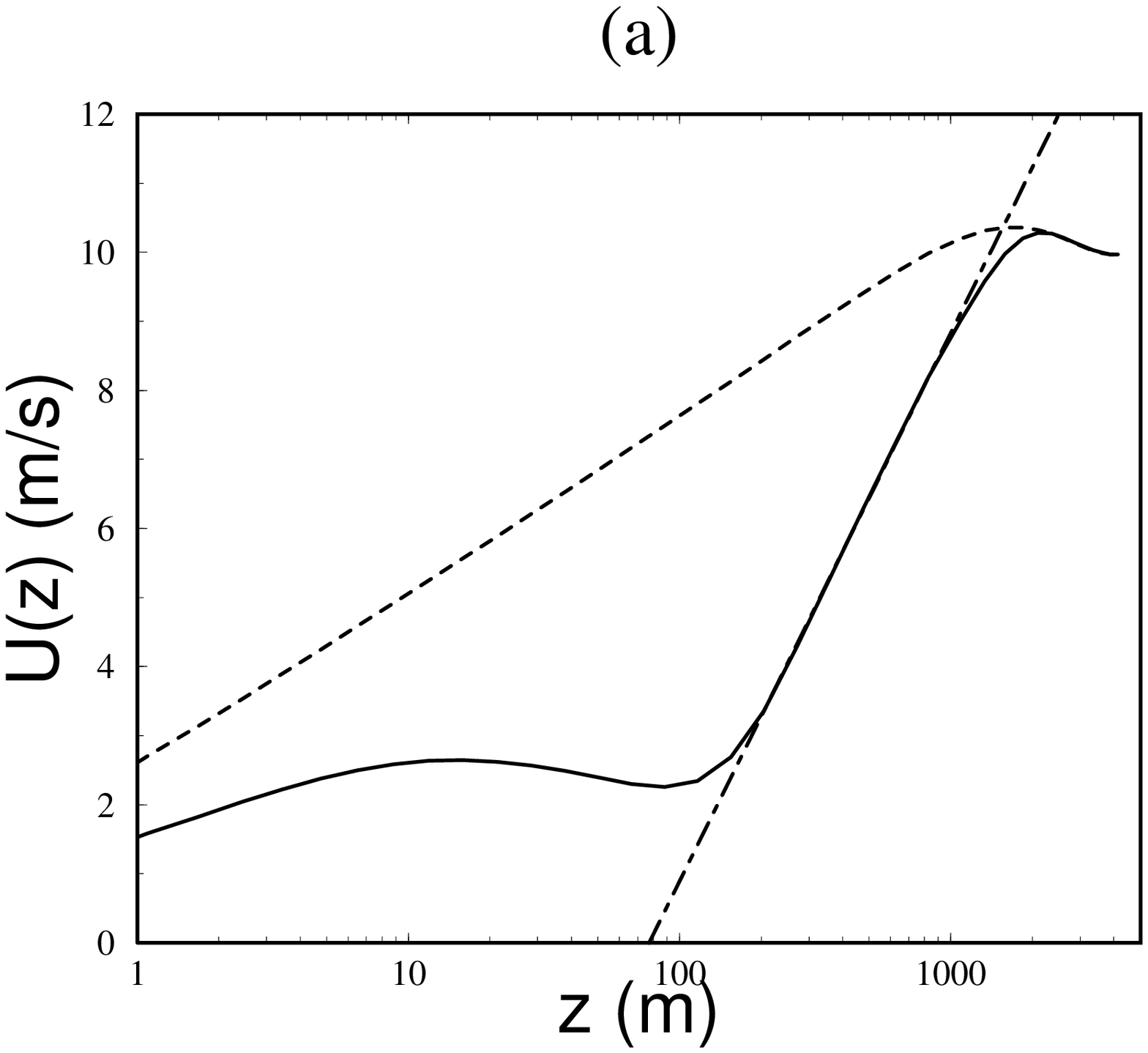,width=.9\linewidth}}
\end{center}
\end{minipage} \hfill
\begin{minipage}{.4\linewidth}
\vspace{0.0cm}
\begin{center}
\mbox{\psfig{figure=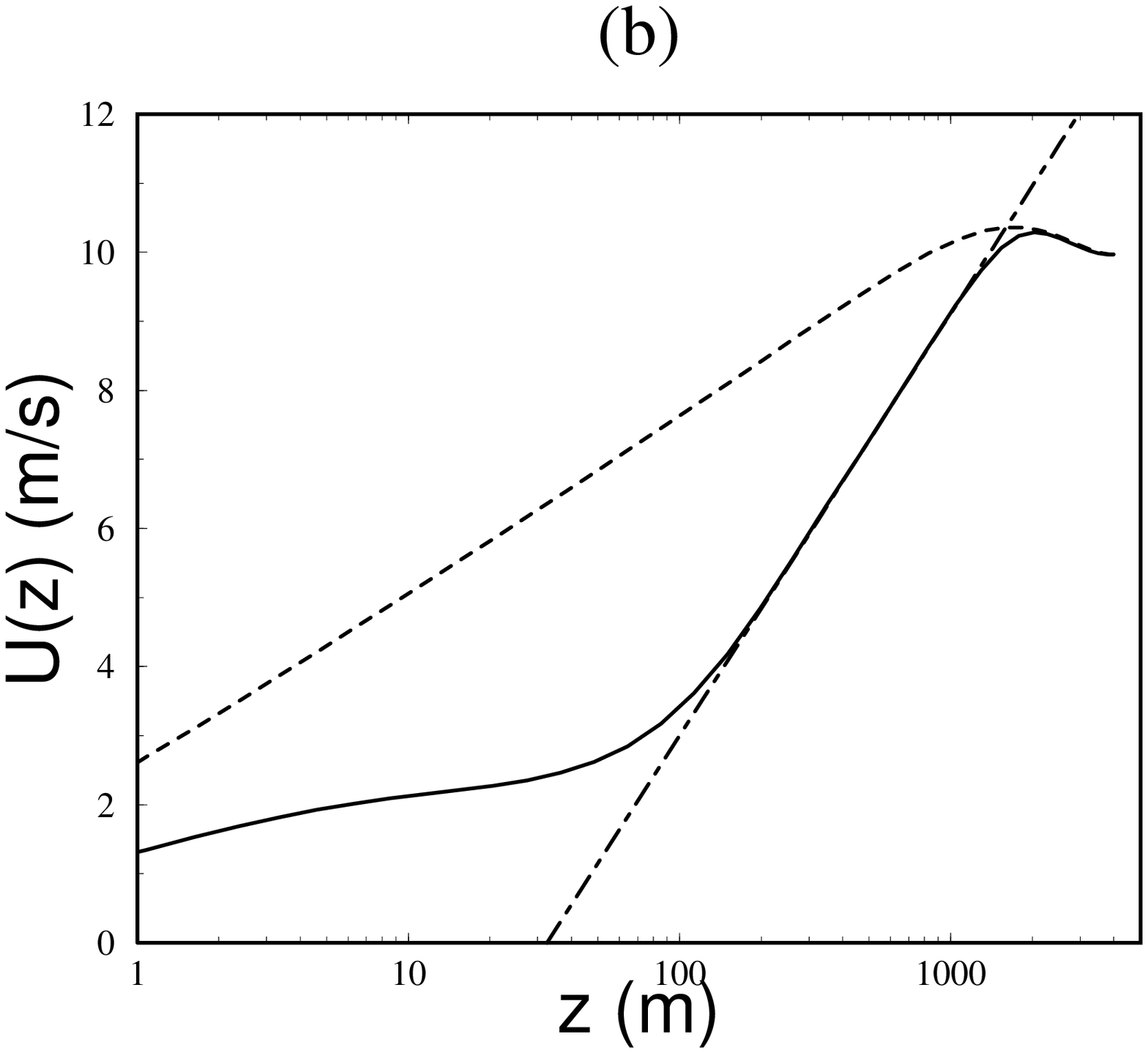,width=.9\linewidth}}
\end{center}
\end{minipage}
\vfill \begin{minipage}{.4\linewidth}
\begin{center}
\vspace{0.0cm}
\mbox{\psfig{figure=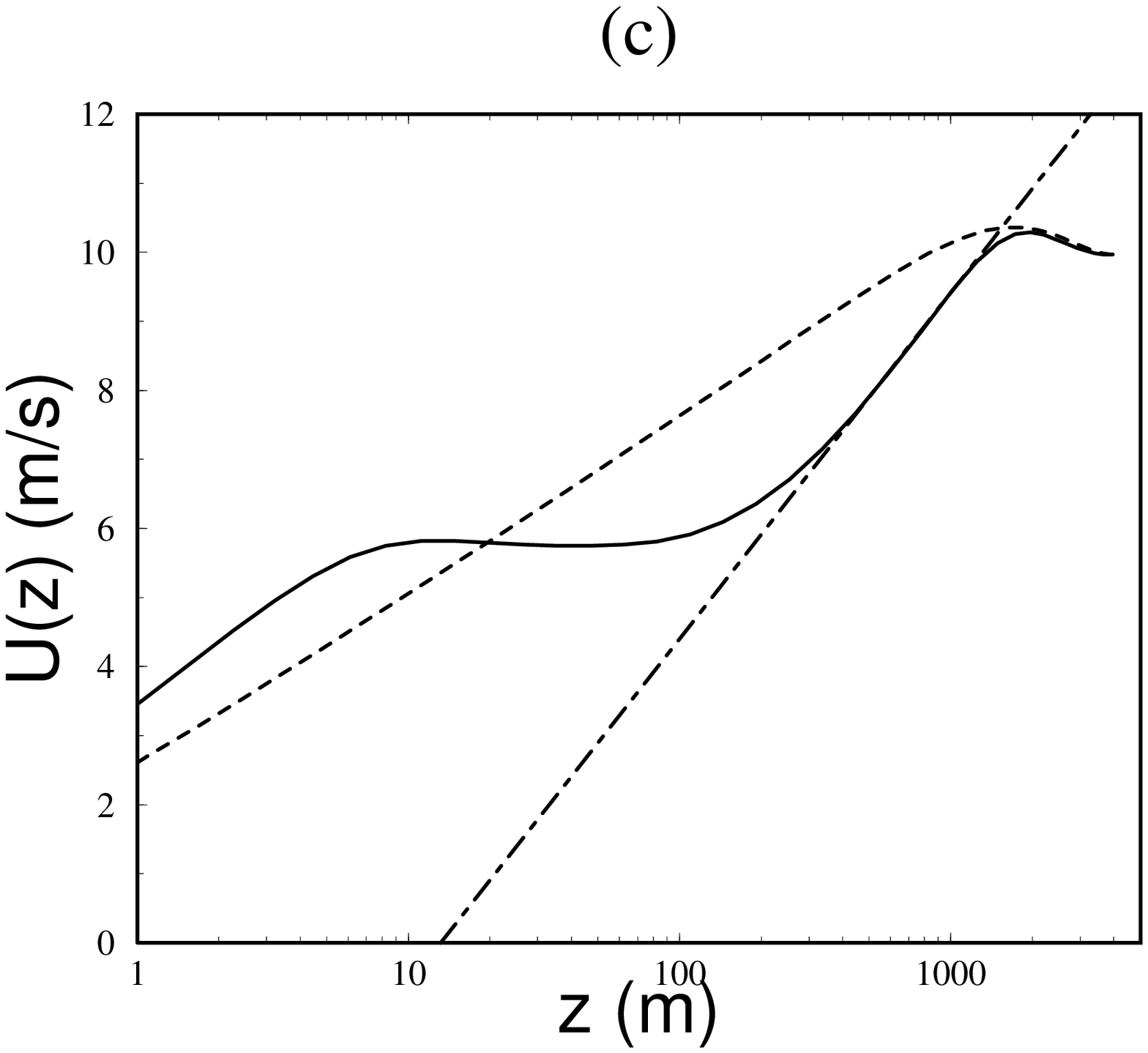,width=.9\linewidth}}
\end{center}
\end{minipage} \hfill
\begin{minipage}{.4\linewidth}
\begin{center}
\mbox{\hspace{3mm}\psfig{figure=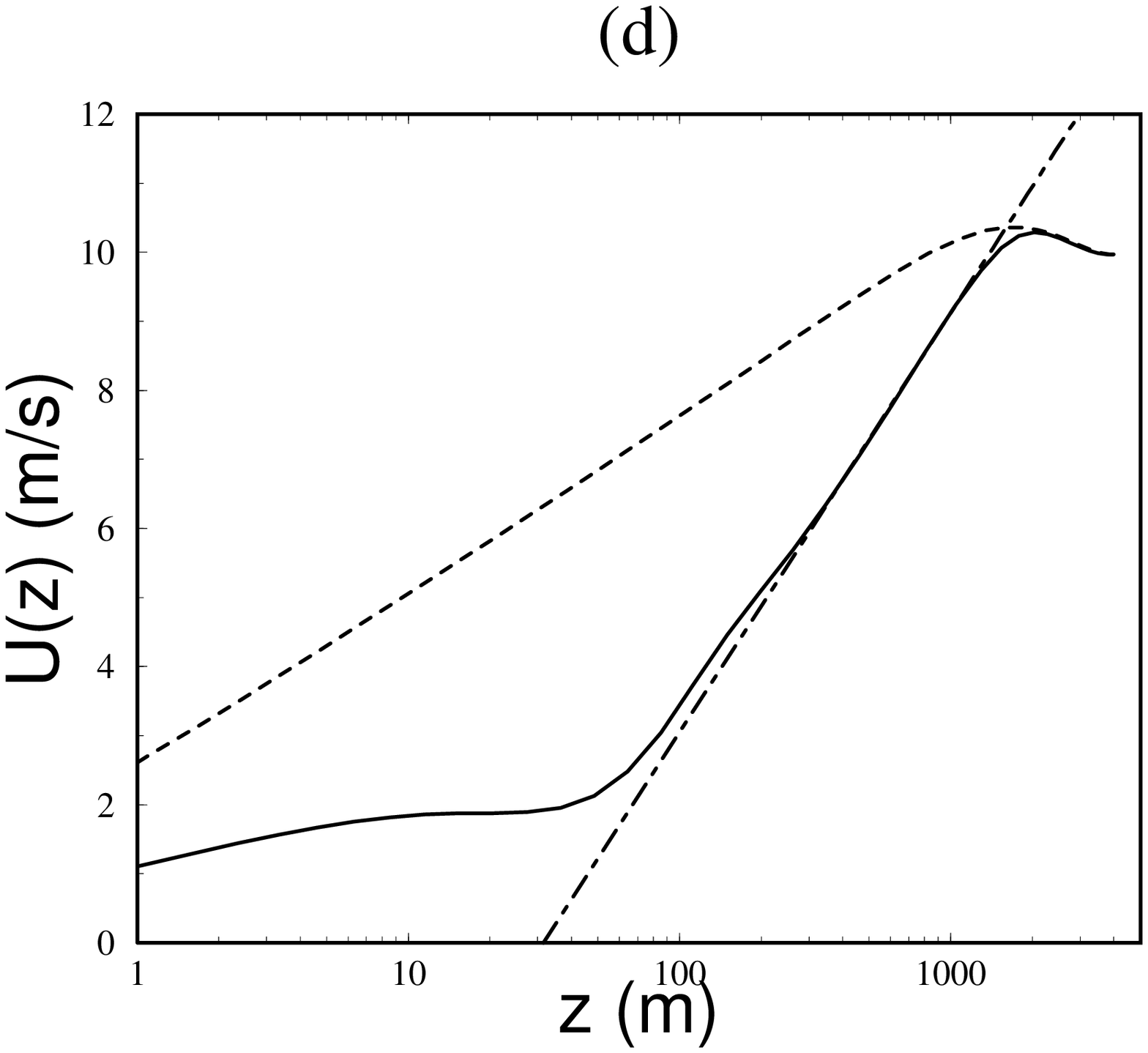,width=.9\linewidth}}
\end{center}
\end{minipage}
\vspace{-0.4cm}
\caption{The local wind speed profiles 
$U(z)=\protect\sqrt{u(z)^2+v(z)^2}$ 
are plotted (solid lines) as
a function of $z$ for four different positions 
($x$ in Eqs.~\protect(\ref{collina}) and \protect(\ref{generalizzo})) 
along the hill H250, corresponding to (a) $x=0$, (b) 
$x=\lambda/4$, (c) $x=\lambda/2$ and (d)
$x=3\lambda/4$.  
No horizontal average has been performed to obtain $U$. 
The dashed lines represent the unperturbed profile. The
dot-dashed lines represent the logarithmic law 
 of Eq.~\protect(\ref{generalizzo}), 
with parameters
$u_{\star}^{\mbox{\tiny eff}}(x)$ and $z_{0}^{\mbox{\tiny eff}}(x)$
obtained by least-square fits performed inside the scaling regions.
The values of these 
 effective parameters are given 
in the text.}
\end{figure}
typical behaviors for the horizontal wind speed profile
(which can be thought to be
representative of an area of the order of the model grid box)
$U(z)=\sqrt{u(z)^2+v(z)^2}$ (for simplicity, 
the dependence on the $x$-coordinate is omitted in the notation)
as a function of $z$ 
are presented in lin-log coordinates for the 
steepest hill H250 and for four values of the $x$-coordinate in 
Eqs.~(\ref{collina}) and (\ref{generalizzo})
corresponding to: (a) $x=0$ 
(i.e. $h=0$), (b) $x=\lambda/4$
(i.e. $h=H/2$ upwind), (c) $x=\lambda/2$ (i.e. $h=H$) 
and (d) $x=3\lambda/4$ (i.e. $h=H/2$ downwind), respectively. 
From this figure, clean scaling regions of the type described by
Eq.~(\ref{generalizzo})
(a straight-line with slope $u_{\star}^{\mbox{\tiny
eff}}(x)/k$ in these coordinates) extended up to $\sim$ 1 decade 
are evident and a reliable measure of both $u_{\star}^{\mbox{\tiny
eff}}(x)$
and $z_0^{\mbox{\tiny eff}}(x)$ is thus feasible by least-square fits.
Specifically, for the four above positions along the hill, we have 
obtained the following values of $u_{\star}^{\mbox{\tiny eff}}(x)$ 
and $z_0^{\mbox{\tiny eff}}(x)$: 1.38 $m/s$, 77.3 $m$ (for $h=0$);
1.07 $m/s$, 32.4 $m$ (for $h=H/2$ upwind);
0.87 $m/s$, 13.2 $m$ (for $h=H$) and 1.05 $m/s$ and 31.5 $m$ 
(for $h=H/2$ downwind), respectively. 
Such values can be 
compared with those in the absence of any hill: $u_{\star}\simeq 0.44\;m/s$
and $z_0\simeq 0.16\;m$. 
Notice that the logarithmic behavior shown in the above figure 
is not peculiar of the four
positions
considered, being indeed
observed for the full range of variability of $x$ and for all the three
hills.

The results of the least-square fits are summarized in Fig.~2
\begin{figure}[ht]

\vfill \begin{minipage}{.4\linewidth}
\begin{center}
\vspace{0.0cm}
\mbox{\psfig{figure=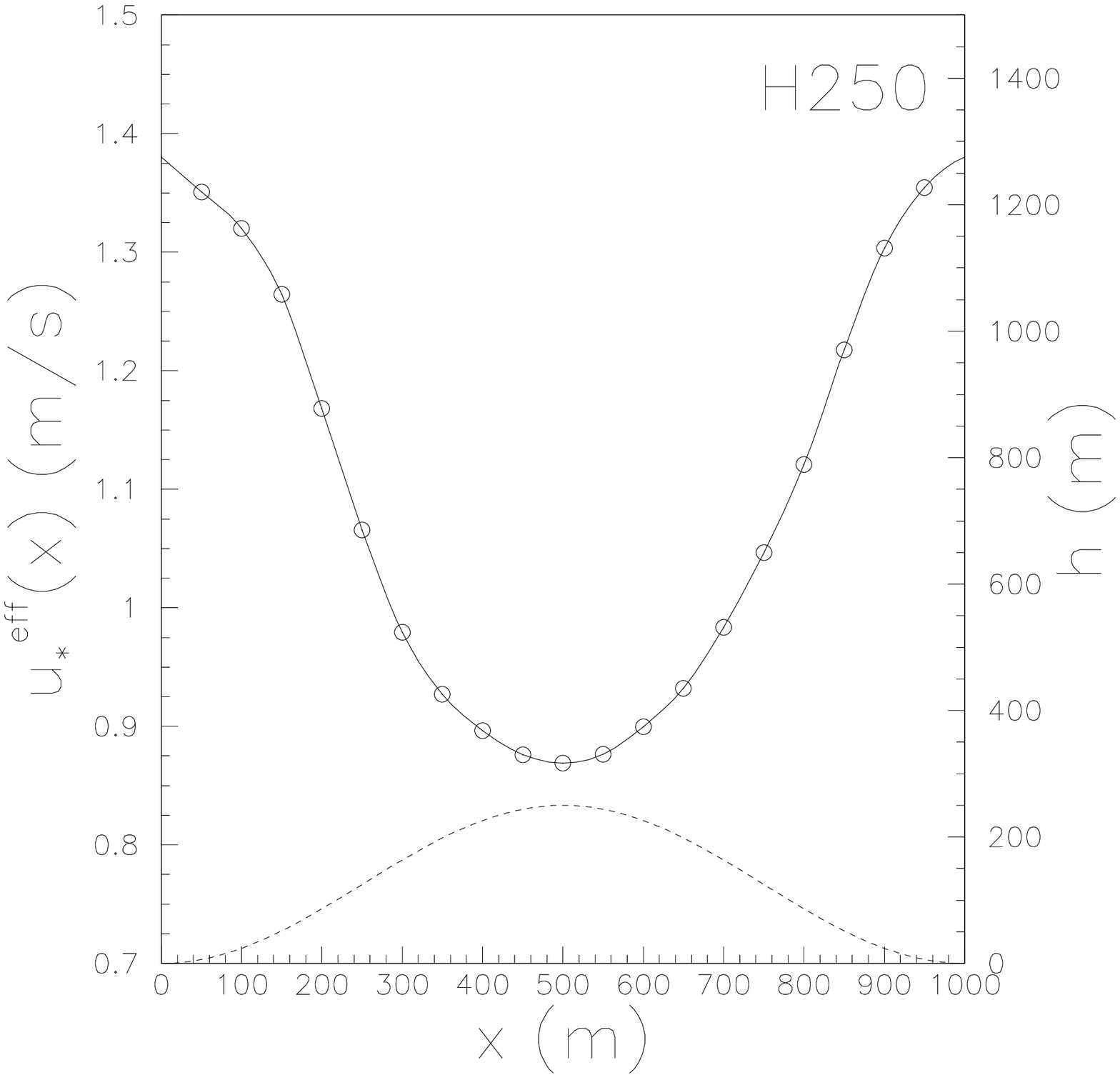,width=.9\linewidth}}
\end{center}
\end{minipage} \hfill
\begin{minipage}{.4\linewidth}
\vspace{0.0cm}
\begin{center}
\mbox{\psfig{figure=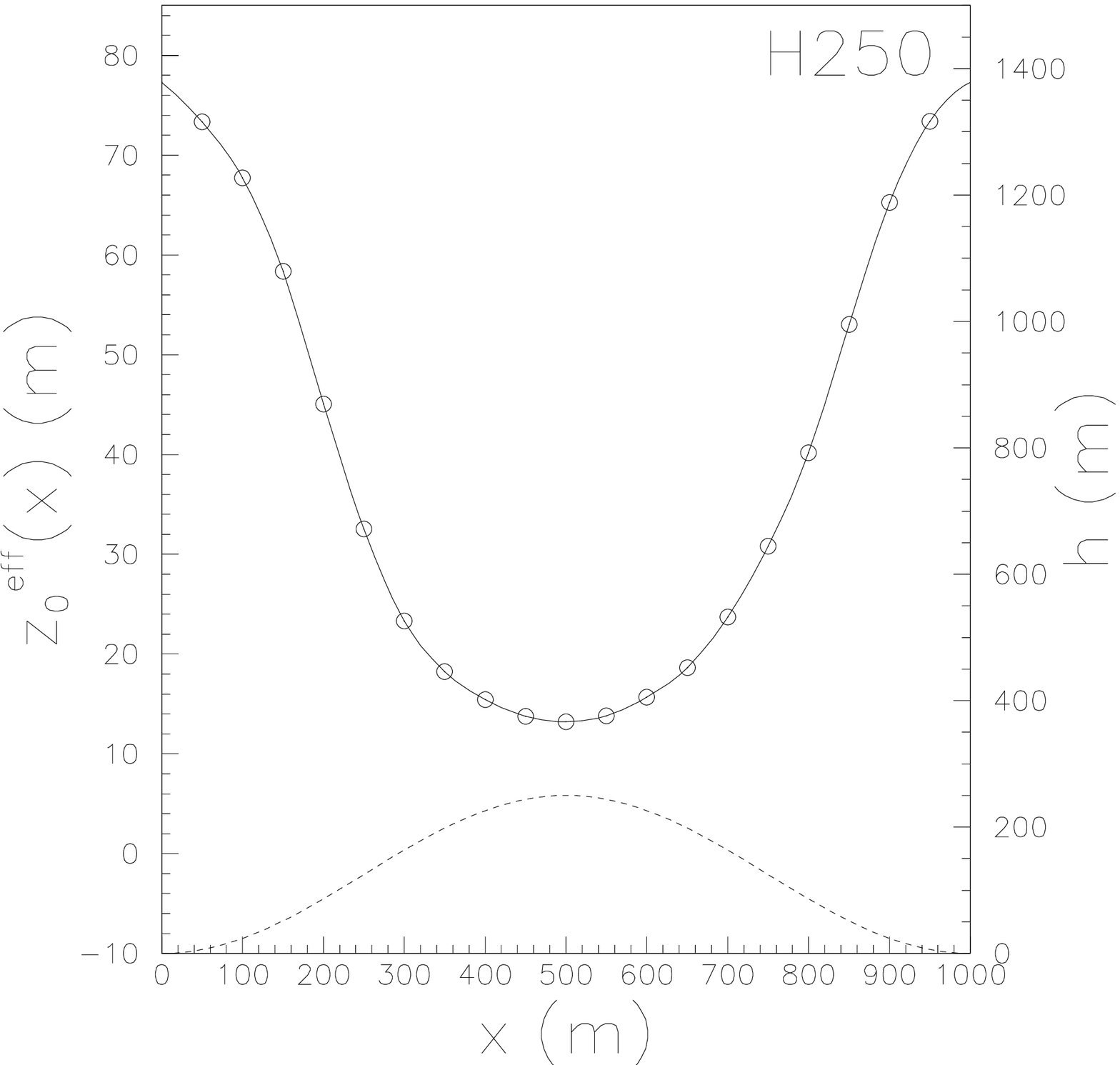,width=.9\linewidth}}
\end{center}
\end{minipage}

\vfill \begin{minipage}{.4\linewidth}
\begin{center}
\vspace{0.0cm}
\mbox{\psfig{figure=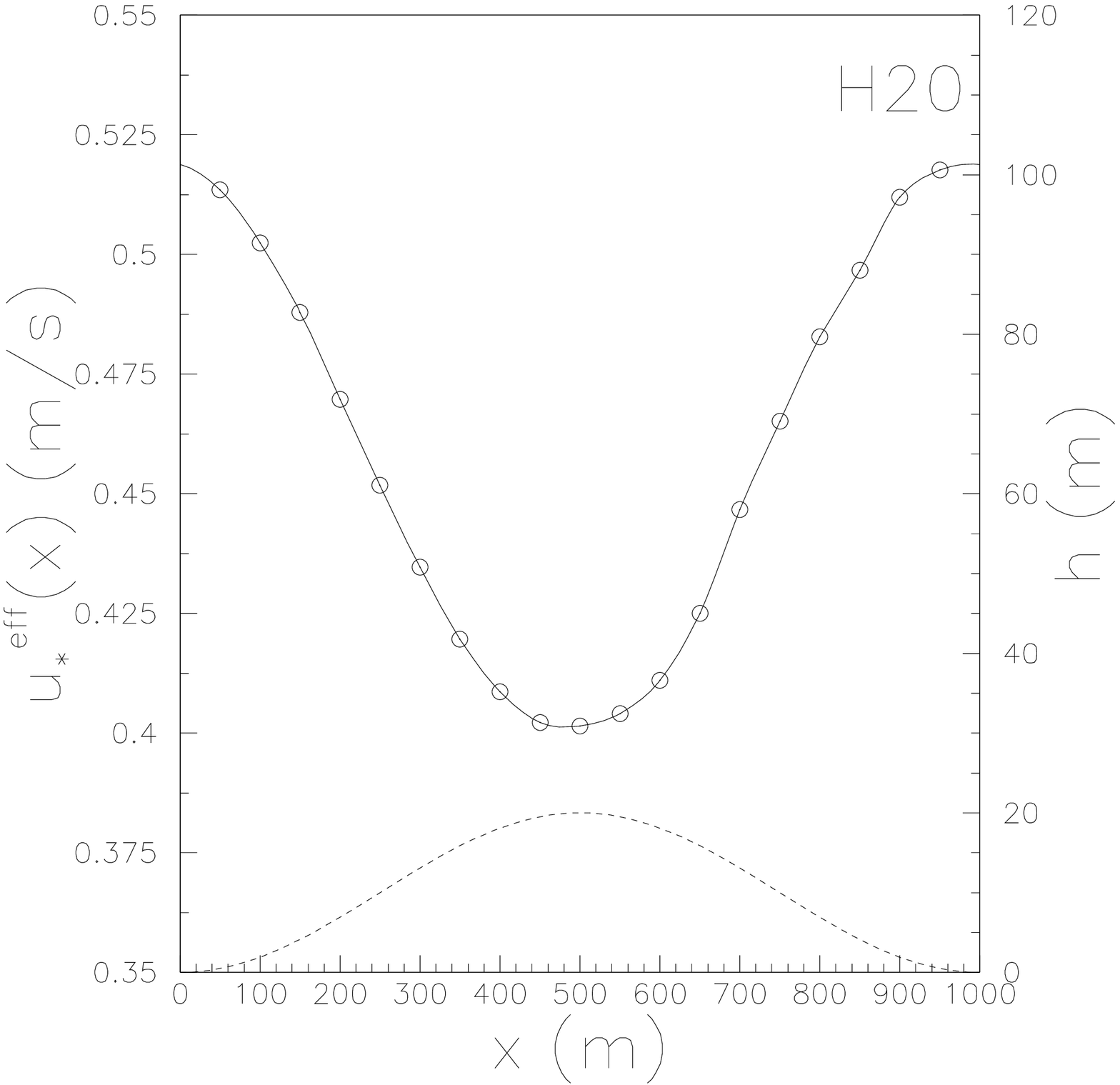,width=.9\linewidth}}
\end{center}
\end{minipage} \hfill
\begin{minipage}{.4\linewidth}
\begin{center}
\mbox{\psfig{figure=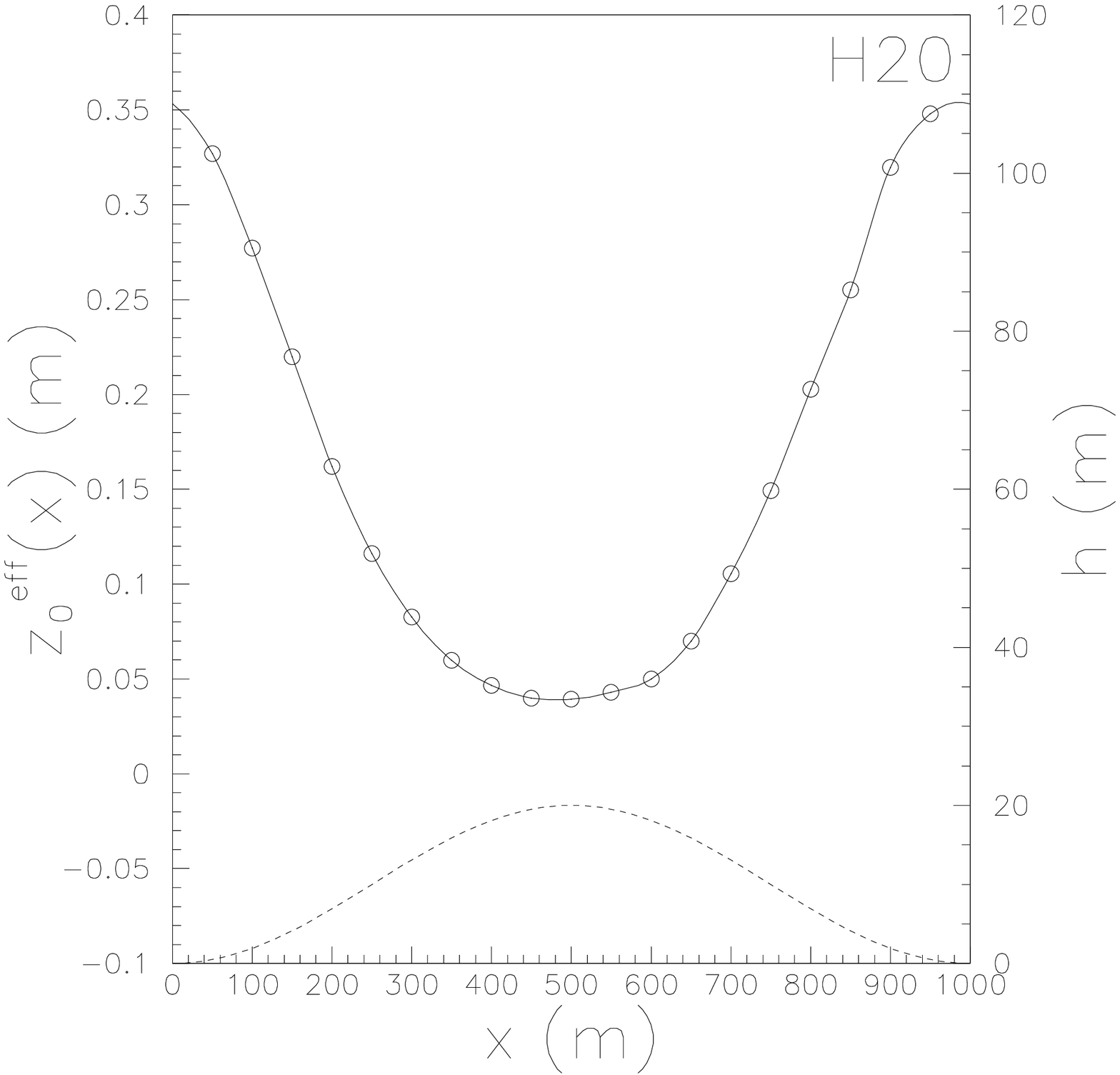,width=.9\linewidth}}
\end{center}
\end{minipage}

\caption{The measured effective parameters 
$u_{\star}^{\mbox{\tiny eff}}(x)$ (on the left) and 
$z_0^{\mbox{\tiny eff}}(x)$ (on the right) (represented with circles
joined by a full line) as a function of the position $x$ along the
axis of the hill H250 (above) and H20 (below). The ordinate
on the right of each plot is relative to the hill elevation (dashed
line).
}
\end{figure}
where both profiles $u_{\star}^{\mbox{\tiny eff}}(x)$ (on the left)
and $z_0^{\mbox{\tiny eff}}(x)$ (on the right) are shown as a function
of $x$ both for the steepest hills (H250) and for the gentlest one (H20) 
(from above to below: 
notice the different scales in the ordinates both on the left and
on the right hand sides).
 Similar behaviors 
have been also observed for the intermediate hill H100.
Notice that both $u_{\star}^{\mbox{\tiny eff}}(x)$
and $z_0^{\mbox{\tiny eff}}(x)$ have a shape similar, but opposite,
to that of the underlying topography, here described by
Eq.~(\ref{collina}).
Furthermore,
we have verified that the use of either the total horizontal speed or just the
$x$-component of the velocity
has little impact on the values of the obtained effective parameters.

Finally, we have observed that both the average values and
amplitudes
of these sinusoidal shapes show, to a reasonable approximation,
a linear dependence on $H/\lambda$, 
with $z_0^{\mbox{\tiny eff}}(x)           \mapsto z_0$ and 
$ u_{\star}^{\mbox{\tiny eff}}(x) \mapsto  u_{\star}$ when 
$ H/\lambda                                    \mapsto 0$.

Notice that the above properties are valid only for 
$z > z_{min}\sim H$, as already written on the right of
Eq.~\protect(\ref{generalizzo}),
while in the underlying flow, nearest to the hill surface, the
upwind-downwind
asymmetry of the velocity field is more pronounced with increasing the 
$ H/\lambda$ ratio (for instance the ``flow separation'' occurring in
the lee
of steep hills and ridges, close to the surface, is a well-known
feature we have observed, for instance, for the hill H250). 

Some other remarks and comments are in order. 
The first concerns the physical meaning of dynamical quantities like 
$u_{\star}^{\mbox{\tiny eff}}(x)$ and $z_0^{\mbox{\tiny eff}}(x)$.
It is natural to assume 
that they share the same meaning of $u_{\star}$ and $z_0$, 
respectively, for flat surfaces but now
with a smooth dependence on the position. 

The minimum of $u_{\star}^{\mbox{\tiny eff}}(x)$
occurring on the top of the hills can be easily understood, being
related to curvature effects.  
If the streamlines are indeed curved  
(as it happens in our case due to the 
presence of the ridges),
energy may be transferred between the large-scale flow and 
the turbulent motion (the amount of which
can be measured as $u_{\star}^{\mbox{\tiny eff}\; 2}(x)$).
This point can be easily grasped by means of the simple argumentations
presented in Ref.~\cite{T80} based on the analysis of the transfer of energy
in terms of the angular momentum of the flow about the axis of curvature.
In this framework, it is possible to show
that the flow curvature above the hill top 
works to transfer the energy from
turbulent motion toward larger scales. This fact reduces the turbulence
energy and thus $u_{\star}^{\mbox{\tiny eff}\;2}(x)$. The situation
changes
above the valleys, where the mechanism is reversed: 
energy is now released from the large-scale flow 
and appears as energy of the turbulent motion, and thus an  
enhancement of $u_{\star}^{\mbox{\tiny eff}\;2}(x)$ occurs.

Also the minimum of $z_0^{\mbox{\tiny eff}}(x)$ 
above the top of the ridge can be understood 
remembering \cite{YM}
that, in the case of flat terrain, $z_0$ is related to
the height of surface protrusions and thus to the 
size of eddies generated by the flow around them. 
Indeed, for hilly terrain it is 
reasonable to assume that the eddies placed in proximity of the
surface act as a sort of effective protrusions for the above
pre-asymptotic flow. In this picture, it is evident that such a
`dynamical
roughness' turns out to be smaller on the hill top, where, 
as already said, turbulence 
(and thus eddies activity) is weaker than on the valleys.

We have already pointed out the presence of an upwind-downwind
symmetry inside the `logarithmic layer'
for both $u_{\star}^{\mbox{\tiny eff}}(x)$ and $z_0^{\mbox{\tiny
eff}}(x)$. Physically, this means that, sufficiently
far above the topography, 
it is just the mean 
cumulative result of many almost independent effects, 
arising in the underlying layer, which is relevant for the large scale 
dynamics. The details of the underlying dynamics 
(e.g., the regime of flow separation)
only affects the numerical 
values of quantities related to the effective parameters (e.g.,
its average values along the hill and the amplitude of the modulation).

We can finally investigate the relation between the asymptotic 
and the pre-asymptotic effective parameters. To that end, averaging
the logarithmic profile (\ref{generalizzo})
over the hill periodicity box, we obviously obtain again the 
log-profile (\ref{logen}), but where now ${\sf u}_{\star}^{\mbox{\tiny eff}}$ 
and ${\sf z}_0^{\mbox{\tiny eff}}$ 
are related to the effective parameters at smaller scales. Specifically,
\begin{equation}
{\sf u}_{\star}^{\mbox{\tiny eff}} = 
\langle u_{\star}^{\mbox{\tiny eff}}(x)\rangle
\qquad\qquad \ln {\sf z}_0^{\mbox{\tiny eff}}= 
\frac{\langle u_{\star}^{\mbox{\tiny eff}}(x)\,
\ln z_0^{\mbox{\tiny eff}}(x)\rangle}{\langle u_{\star}^{\mbox{\tiny 
eff}}(x)\rangle}\;\;\; ,
\label{relatio}
\end{equation}
which do not depend on the position along the hill. The results thus
remain unchanged when the average is performed over scales larger than
the hill periodicity box.\\
The relationships between the values of these parameters
and the topography characteristics are discussed, for instance, in
Refs.~\cite{W92,WM93}.
It is easily checked that the first of the two above relations is an 
immediate consequence of the conservation of the total flux of vertical 
momentum inside the whole domain. More interestingly,
the second relation tells us that the roughness parameter 
${\sf z}_0^{\mbox{\tiny eff}}$ does not have solely a geometrical
meaning, 
being in fact related to $u_{\star}^{\mbox{\tiny eff}}(x)$ at smaller
scales.
This fact places in an unfavorable light attempts to evaluate, 
either experimentally or numerically, the parameter 
${\sf z}_0^{\mbox{\tiny eff}}$
without taking into account its dependence on the flow 
configuration as a whole. 

We stress that the possibility of describing the flow in terms
of effective parameters is a direct consequence of an intrinsic
scale separation in the dynamics.
There is a simple (and more treatable)
physical system where this link between scale separation and effective
parameters can be rigorously stated. This is
the transport of passive scalar field (e.g., the density of particles 
or dye injected into the flow). In this problem,
if a small-scale turbulent velocity field
${\bf v}({\bf x},t)$ (varying on scales of the order of $l_0$) is superimposed
to an uniform field ${\bf V}$,  
for times large compared with 
those characteristic of the turbulent field,
the concentration field  $\Theta$ 
is now varying on scales $L>>l_0$ and, as shown in Ref.~\cite{BCVV95},
obeys the Smoluchowsky equation:
\begin{equation}
\partial_t\Theta +({\bf V}\cdot{\bf
\partial})\,\Theta={\bf\partial}(D\,{\bf\partial})\Theta
\label{FP}
\end{equation}
where $D$ is the so-called eddy-diffusivity, i.e. an effective enhanced 
diffusivity which  depends
on the characteristics of turbulence as a whole. 

Furthermore, under the assumption that ${\bf V}$ is now varying
on spatial scales of the order of  $L>>l_0 $, 
it has been shown (see Ref.~\cite{M97}) that again 
a Smoluchowsky equation for $\Theta$ 
holds for large times, but where the
effective diffusivity  now takes a smooth dependence 
on the position on scales of the order of $L$. 

The existence of some analogy between the two aforesaid contexts
seems clear. The role
played by ${\bf v}({\bf x},t)$ in the passive scalar dynamics is played
in our problem by the smallest eddies placed in the vicinity 
of the wall and produced by the roughness elements (flat surface)
or by the topography (flow averaged over distances much larger than the 
topography wave-length). 
The statistical role of such small scale features affects the
large scale dynamics 
only via parameters such as $D$ for the passive scalar problem
and such as $u_{\star}$ and $z_0$ in the case of the flow above a flat
surface
or ${\sf u}_{\star}^{\mbox{\tiny eff}}$ and
${\sf z}_0^{\mbox{\tiny eff}}$  
in the case of the flow over hilly terrain, averaged at a scale much 
larger than the topography wave-length (asymptotic regime). 
On the other hand, the passive
scalar dynamics observed at the same scales (of the order of $L$)
of the varying advecting
velocity, ${\bf V}$, corresponds in our problem to the velocity 
behavior  above a hilly terrain, when this velocity field is observed at
scales comparable with the integral scale 
of the problem (the wave-length, $\lambda$, of the hill). In such
(pre-asymptotic)  regimes, the dynamics is again described in terms 
of effective parameters but now depending on the position, i.e.
$u_{\star}^{\mbox{\tiny eff}}(x)$ and $z_0^{\mbox{\tiny eff}}(x)$ 
in the case of flow above a hilly surface, 
$D({\bf x})$ for the passive scalar problem.

The analogy cannot be pushed any further, 
but we are convinced that it is a clear indication that the scenario we
have described is a direct  consequence of an intrinsic scale separation 
in the dynamics.

In conclusion, we have presented strong numerical evidences
showing the `local' validity of the law-of-the-wall 
not only above flat terrain, where it is well known, but also 
in the presence of hilly terrain. 
The logarithmic shape for the velocity field
is restored far enough from the terrain  where
typical scales of the velocity fields appear ``separated'' from those 
relative to
the smallest eddies confined near the surface.
Local logarithmic profiles are well evident and
described in terms of effective
parameters showing a smooth dependence on the position along the hill.
We have found \cite{BMR00}
(but not discussed in this Letter) similar
results analyzing data from 
both wind tunnel experiments (see Ref.~\cite{GTD96})
and experiments in nature (see Ref.~\cite{Hignett94}). Such results confirm the
scenario here outlined and also give a strong evidence that our 
conclusions do not depend on the particular choice of the
parameterization scheme adopted in the numerical model here considered.\\
The relation between asymptotic and pre-asymptotic 
effective parameters
is also derived and the dynamical role of the roughness clearly 
emerges.
Finally, a simple analogy with the pre-asymptotic dynamics 
for the passive scalar problem,
is exploited to confirm our interpretation of the reasons
underlying the presented  numerical evidence.

\vskip 0.2cm
{\bf Acknowledgements}
We are particularly grateful to N.~Wood for 
providing us with his data-set relative to velocity fields as well as 
many useful comments. Helpful discussions and suggestions by
D.~Anfossi, R.~Festa, E.~Fedorovich, S.~Gallino, D.~Mironov, 
S.~Nazarenko, G.~Solari, 
F.~Tampieri and P.A.~Taylor are also acknowledged.
This work was partially supported by the INFM PA project GEPAIGG01.


\begin{thebibliography}{99}

\bibitem{T86} S.~Tibaldi, 
{\em Adv. Geophys.}, {\bf 99}, 341 (1986).

\bibitem{W92} N.~Wood,
Turbulent flow over three-dimensional hills,
(Ph.D. dissertation, University of Reading, UK., pp. 223, 1992).

\bibitem{T81} P.A.~Taylor, 
{\em Q.J.R. Meteorol. Soc.}, {\bf 107}, 111 (1981). 

\bibitem{E87} S.~Emeis,
{\em Bound. Layer Meteorol.}, {\bf 39}, 379 (1987).


\bibitem{TSM89} P.A.~Taylor, R.I.~Sykes and P.J.~Mason,
{\em Bound. Layer Meteorol.}, {\bf 48}, 409 (1989).

\bibitem{WM93} N.~Wood and P.~Mason,
{\em Q.J.R. Meteorol Soc.}, {\bf 119}, 1233 (1993).

\bibitem{EZ94} S.~Emeis and S.S.~Zilitinkevich,
{\em Bound. Layer Meteorol.}, {\bf 55}, 191 (1994).

\bibitem{YM}
A.S.~Monin and A.M.~Yaglom, {\it Statistical Fluid Mechanics},
(MIT Press, Cambridge, Mass., 1975).

\bibitem{N99}
S.~Nazarenko, ``On exact solutions for near-wall turbulence theory''.
{\em Phys. Rev. Lett} (submitted), 2000.

\bibitem{NKD99}
S.~Nazarenko, N.K.-R.~Kevlahan and B.~Dubrulle,
``Nonlinear RDT theory of near-wall turbulence'',
{\em Physica D} (submitted), 2000.


\bibitem{GTD96}
W.~Gong, P.~Taylor and A.~Dornbrack,
{\em J. Fluid Mech.}, {\bf 312}, 1 (1996).

\bibitem{Hignett94} P.~Hignett and W.P.~Hopwood,
{\em Bound. Layer Meteorol.}, {\bf 68}, 51-73 (1994).

\bibitem{H79}
              J.R.\ Holton,
              {\it An introduction to dynamic meteorology}
              (Academic Press, London, 1979).

\bibitem{T80}
              A.A.\ Townsend,
              {\it The structure of turbulent shear flow}
              (Cambridge University Press, Cambridge, 1980).


\bibitem{BCVV95}
L.~Biferale, A.~Crisanti, M.~Vergassola and A.~Vulpiani,
{\em Phys. Fluids}, {\bf 7}, 2725 (1995).

\bibitem{M97}
A.~Mazzino,
{\em Phys. Rev. E}, {\bf 56}, 5500 (1997).

\bibitem{BMR00}
S.~Besio, A.~Mazzino and C.F.~Ratto, 
``Local law-of-the-wall in complex topography: 
a confirmation from wind tunnel experiments'', in preparation.

\end{thebibliography}
\end{document}